\documentclass[aps,pra,twocolumn,floatfix,superscriptaddress]{revtex4-1}
\usepackage{subfigure,dcolumn}
\usepackage[T2A,T1]{fontenc}
\usepackage[english]{babel}
\usepackage{braket}
\usepackage{graphicx}
\usepackage[colorinlistoftodos]{todonotes}
\usepackage[utf8]{inputenc}
\usepackage{amsthm}
\usepackage{amsmath}
\usepackage{amsfonts}
\usepackage{slashed}
\usepackage{hyperref}
\usepackage[normalem]{ulem}
\usepackage{soul}

\begin{document}

\title{On the limits of photon-mediated interactions in one-dimensional photonic baths}

\author{Eduardo Sánchez-Burillo}
\affiliation{Max-Planck-Institut für Quantenoptik, D-85748 Garching, Germany}
\author{Diego Porras}
\affiliation{Instituto de Física Fundamental IFF-CSIC, Calle Serrano 113b, E-28006 Madrid, Spain}
\author{Alejandro González-Tudela}
\affiliation{Instituto de Física Fundamental IFF-CSIC, Calle Serrano 113b, E-28006 Madrid, Spain}
\email{a.gonzalez.tudela@csic.es}

\date{\today}

\begin{abstract}
	
The exchange of off-resonant propagating photons between distant quantum emitters induces coherent interactions among them. The range of such interactions, and whether they are accompanied by dissipation, depends on the photonic energy dispersion, its dimensionality, and/or the light-matter couplings. In this manuscript, we characterize the limits of photon-mediated interactions for the case of generic one-dimensional photonic baths under the typical assumptions, that are, having finite range hoppings for the photonic bath plus local and rotating-wave light-matter couplings. In that case, we show how, irrespective of the system's parameter, the coherent photon-mediated interactions can always be written as a finite sum of exponentials, and thus can not display a power-law asymptotic scaling. As an outlook, we show how by relaxing some of these conditions, e.g., going beyond local light-matter couplings (e.g., giant atoms) or with longer-range photon hopping models, power-law interactions can be obtained within certain distance windows, or even in the asymptotic regime for the latter case.

\end{abstract}



\maketitle

\section{Introduction}

Even if perfectly isolated, distant quantum emitters can interact through the fluctuations of the electromagnetic field (photons) around them~\cite{lehmberg70a,lehmberg70b}. The exchange of off-resonant photons, that are the ones with energies different from the emitter's transition frequency, leads to reversible excitation transfer between the emitters because photons are only \emph{virtually} populated during the exchange process. In free-space~\cite{lehmberg70a,lehmberg70b}, for example, these interactions ($J_{ij}$) decay with the distance between emitters ($r_{ij}$) as a power-law $r_{ij}^{-3 (1)}$ in the near (far) field. Such power-law coherent interactions have raised a lot of interest because they can be harnessed for quantum information~\cite{saffman10a,hammerer10a} or simulation tasks, e.g., to explore long-range interacting spin models that are known to lead to many unconventional phenomena~\cite{porras04a,kim10a,hauke10c,sandvik10a,maik12a,islam13a,hauke13a,juneman13a,gong14a,fossfeig15a,koffel12a,kastner11a,vodola14a,gong16a,gong16b,nevado16a,eldredge17a,gong17a,maghrebi17a,zunkovic18}. Unfortunately, in free space these dipolar interactions are accompanied by collective (and individual) dissipative couplings ($\gamma_{ij}$) induced by the resonant photons, precluding many of their potential applications. 

A way of avoiding this problem consists in modifying the photonic environment around the emitters' to inhibit the modes around the emitter's transition frequencies~\cite{purcell46a}. This can be done, for example, in photonic crystals~\cite{joannopoulos97a,bykov75a, kurizki90a,john90a}, where one can indeed cancel the associated dissipation, i.e., $\gamma_{ij}=0$, by tuning the emitter's frequency into a photonic band-gap region. This cancellation generally comes at the price, however, of an exponential localization of the interactions, i.e., $J_{ij}\propto e^{-r_{ij}/\xi}$, whose characteristic length $\xi$ can be tuned by changing the band energy dispersion and the emitter's detuning to the band-edge~\cite{douglas15a,gonzaleztudela15c}. The only exceptions to this exponential localization of such photon-mediated interactions, to our knowledge, have been found in high-dimensional singular band-gaps~\cite{gonzaleztudela18c,perczel18a,gonzaleztudela18d,garciaelcano19a,ying19a}, where power-law $J_{ij}$ interactions have been predicted with no associated dissipation. While some attempts with similar energy dispersions have been explored in 1D~\cite{sanchezburillo19a}, the emergence of such power-law coherent interactions mediated by one-dimensional photonic environments remains so far elusive. 

Motivated by this quest, in this manuscript we study the limits in the range of photon-mediated interactions induced by one-dimensional environments. Given the variety of experimental platforms available nowadays to explore such quantum optical effects, ranging from photonic crystals~\cite{goban13a,lodahl15a}, circuit QED metamaterials~\cite{liu17a,mirhosseini18a}, subwavelength atomic arrays~\cite{rui20a,masson19a}, or state-dependent optical lattices~\cite{devega08a,krinner18a}, we provide results for generic one-dimensional models using a minimal set of assumptions, that are, having local and excitation-conserving light-matter couplings, together with finite range hoppings for the bath. With these assumptions, we are able to show that the quantum emitters' interactions can always be written as a finite sum of exponential terms and can thus never display a power-law decay irrespective of the model considered. Besides, we also study situations where some of these assumptions are broken, and show how one could obtain (quasi) power-law interactions. For example, we show that non-local light-matter couplings, as the ones enabled by giant atoms~\cite{kockum19a,kannan19a}, open up the possibility of mimicking power-law interactions up to certain distances in a controlled way. Furthermore, we also study baths with longer-range hoppings, which can lead to power-law interactions even in the asymptotic limit. 

The manuscript is structured as follows: in Sec.~\ref{sect:model} we first write down the generic light-matter Hamiltonian that we will consider along the manuscript. Then, we derive the effective photon-mediated interactions in Sec.~\ref{sect:eff_int}. Afterwards, in Sec.~\ref{sect:power_law} we explore the possibilities of obtaining power-law interactions by breaking some of the assumptions of the general model considered in Sec.~\ref{sect:model}. Finally, we summarize our findings in Sec.~\ref{sect:conclusions}.

\begin{figure}[tbh!]
    \centering
    \includegraphics[width=0.95\columnwidth]{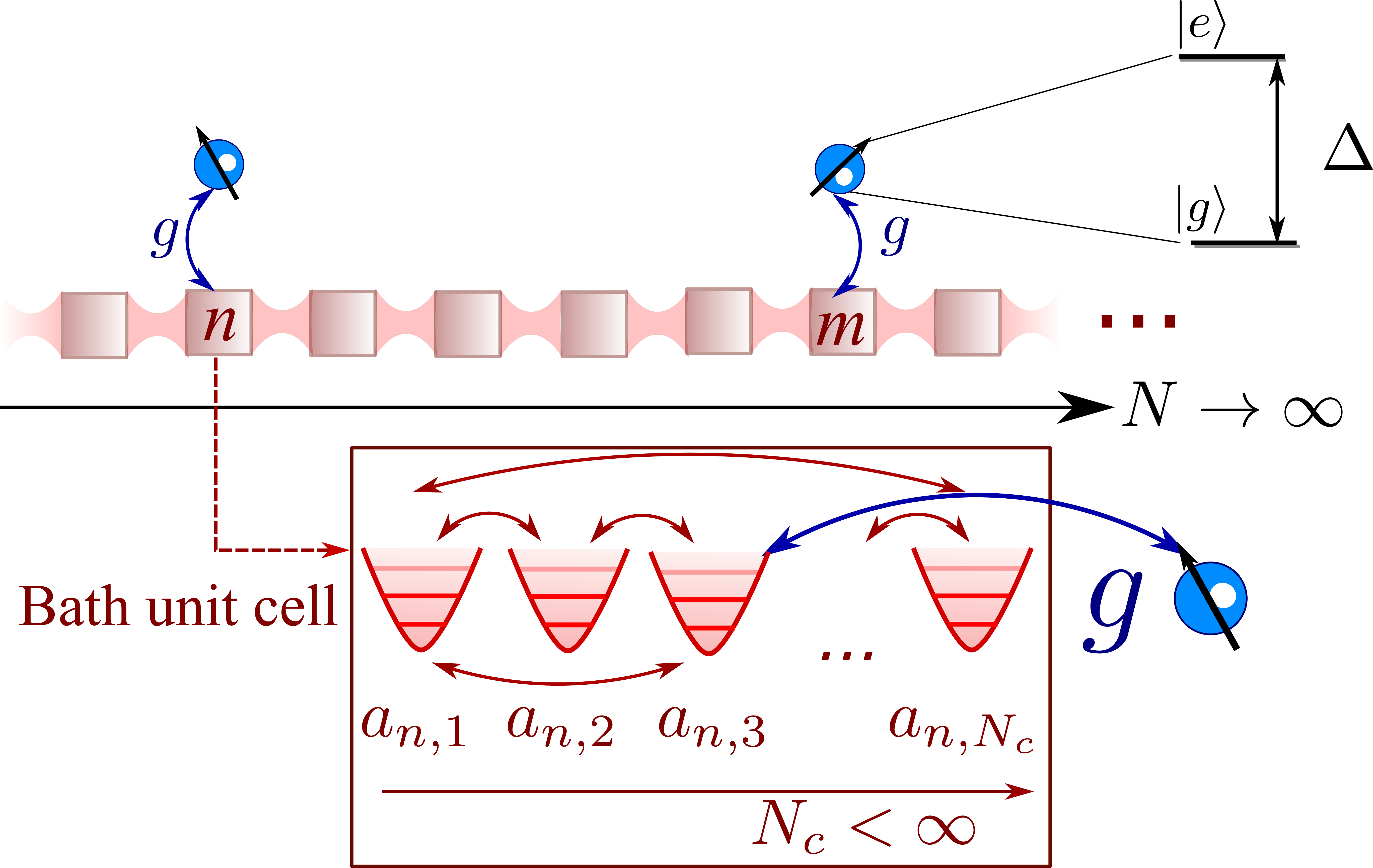}
    \caption{Scheme of model considered: The photonic bath is described by a set of $N$ unit cells (that we will consider to be infinite) composed of $N_c$ (finite) coupled resonator modes. We use bosonic operators $a^\dagger_{n, \alpha}$ to describe the $\alpha$ resonator mode of the $n$-th unit cell. We consider the energy of the discrete resonators $a_{n,\alpha}$ to be equal, and take it as the energy reference of the problem. Two-level ($\ket{g},\ket{e}$) quantum emitters, with detuning $\Delta$ from the energy of the resonators, are locally coupled with strength $g$ to a particular resonator of a given unit cell $n$.}
    \label{fig:sketch}
\end{figure}

\section{Model}\label{sect:model}

We introduce here the different terms of the Hamiltonian of the general model that we will consider along the manuscript (see Fig.~\ref{fig:sketch}). The photonic bath is described by $N$ unit cells with $N_c$ discrete coupled resonators each, that is, a one-dimensional model with $N_c$ sublattices. We only allow for hoppings within the same unit cell or between nearest-neighboring unit cells. Thus, we can capture any one-dimensional photonic bath with finite-range hoppings up to $N_c$ neighbours.  We use bosonic operators $a_{n,\alpha}, a_{n,\alpha}^\dagger$ to describe the photonic excitations of the $\alpha$-th resonator within the $n$-th unit cell of the lattice. As we are interested in the limit $N\rightarrow \infty$, we can safely take periodic boundary conditions and write the bath Hamiltonian in the following form:
\begin{equation}\label{eq:HB}
H_B = \sum_k (\tilde{a}_{k,1}^\dagger \dots \hat{a}_{k,N_c}^\dagger )h_B(k)
\left(
\begin{array}{c}
\hat{a}_{k,1}  \\
\vdots  \\
\hat{a}_{k,N_c}
\end{array}
\right),
\end{equation}
where we use the $\hat{.}$-notation to distinguish the bosonic operators defined in momentum space as: $\hat{a}_{k,\alpha}=\frac{1}{\sqrt{N}}\sum_{n=1}^N e^{-i k n} a_{n,\alpha}$, with $\alpha=1,\dots,N_c$~\footnote{Note that by taking this definition, we are assuming that the unit of distance will be given by the lattice constant. Thus, from now on all the lengths (and momenta) will be units of the lattice constant (or its inverse).}. The matrix $h_B(k)$ is an Hermitian matrix which can be written with full generality as:
\begin{equation}\label{eq:hBk}
h_B(k) = \left(\begin{array}{ccccc}
\delta_1(k) & f_{12}(k) & f_{13}(k) & \dots & f_{1N}(k) \\
f_{12}^*(k) & \delta_2(k)  &  f_{23}(k) & \dots & f_{2N}(k) \\
f_{13}^*(k) & f_{23}^*(k) & \delta_3(k) & \dots & f_{3N}(k) \\
\dots & \dots & \dots &\dots & \dots \\
f_{1N}^*(k) & f_{2N}^*(k) & f_{3N}^*(k) & \dots & \delta_N(k)
\end{array}\right).
\end{equation}
where  $\delta_{\alpha}(k)$ is the dispersion relation of the $\alpha$-th sublattice and $f_{\alpha,\beta}(k)$ the function characterizing the interactions between the $\alpha$-th and $\beta$-th sublattices. This bath Hamiltonian can be diagonalized resulting in $N_c$ energy bands $\omega_\alpha(k)$.

The $N_e$ quantum emitters are described as two-level systems ($\ket{g},\ket{e}$), with energies $\Delta$ (see Fig.~\ref{fig:sketch}). Thus, their internal dynamics is simply given by:
\begin{equation}\label{eq:H}
H_S= \Delta \sum_{j=1}^{N_e} \sigma_j^\dagger \sigma_j\,,
\end{equation}
with $\sigma_j=\ket{g}_j\bra{e}$ being the spin-operator transition of the $j$-th atom.
As mentioned in the introduction, we consider that these emitters are locally coupled to the environment (local-dipole approximation).  This implies that each quantum emitter couples only to one of the resonators of a given unit cell of the lattice. Besides, we describe the light-matter interaction through a Jaynes-Cummings Hamiltonian (rotating-wave regime approximation). The latter is a good description of light-matter interaction as long as the coupling strength is much smaller than the emitter/bath frequencies~\cite{cohenbook92a}, as it is typical case in most of the systems of interest to us. Under these assumptions the light-matter Hamiltonian reads:
\begin{equation}\label{eq:Hint}
H_\text{int} = \sum_{j=1}^{N_e} \left(g\sigma_j^\dagger a_{n_j,\alpha_j} + \text{H.c.}\right)\,,
\end{equation}
where $(n_j,\alpha_j)$ denotes the indices of the unit cell $n_j$ and the particular resonator $\alpha_j$ that the $j$-th emitter is coupled to. Summing up, the global generic Hamiltonian that we will consider contains the sum of the three terms: 
\begin{equation}
H=H_S+H_B+H_\mathrm{int}\,,\label{eq:globalH}
\end{equation}

\section{Photon-mediated interactions}\label{sect:eff_int}

In order to obtain the photon-mediated interactions emerging in this general class of models defined by $H$, we will assume to be in the Born-Markov regime in which the photonic bath timescales are much faster than the induced emitter ones. This allows us to adiabatically eliminate the photons~\cite{cohenbook92a,gardiner_book00a}, resulting in photon-mediated interactions containing both a real and an imaginary part:
\begin{align}
J_{ij}-i\gamma_{ij}=\sum_E \frac{\bra{0}\sigma_j H_\text{int}\ket{E}\bra{E}H_\text{int}\sigma_i^\dagger\ket{0}}{\Delta+i0^+-E},\label{eq:Jij_E}
\end{align}
which leads to unitary/non-unitary emitter dynamics, respectively. Here, $\ket{E}$ is an eigenstate of the free part of the Hamiltonian (Eq.~\eqref{eq:globalH} with $H_\text{int}=0$), $E$ is its energy, and $\ket{0}$ is the global vacuum state of the system $\hat{a}_{k,\alpha}\ket{0}=\sigma_j\ket{0}=0$ for all $k$, $\alpha$, and $j$. 

In this manuscript, we are interested only in the situations where $\gamma_{ij}\equiv 0$ which can be obtained within this approximation assuming that $\Delta$ lies in a band-gap region of the model, i.e., $\Delta\notin \omega_\alpha(k)$ for any $\alpha$ or $k$. Note, that this regime can always be obtained in these models since we are considering finite bath hoppings, which impose a finite bandwidth for the energy bands of the model $\omega_\alpha(k)$. Tuning the emitters into those frequency regions, then the photon-mediated interactions result in an effective spin model:
\begin{align}
H_\mathrm{eff}=\sum_{i,j}\left(J_{ij}\sigma_i^\dagger \sigma_{j}+\mathrm{H.c.}\right)\,,
\end{align}
where $J_{ij}$ can be written as (see Appendix):
\begin{equation}\label{eq:Jij}
J_{ij} = \frac{|g|^2}{2\pi}\int_{-\pi}^\pi dk (\Delta\mathbb{I}-h_B(k))^{-1}_{\alpha_i \alpha_j} e^{ikn_{ji}},
\end{equation}
with $n_{ji}\equiv n_{j}-n_{i}$ being the inter-emitter distance, and $\mathbb{I}$ the identity matrix. The integrand can be expanded as:
\begin{align}
((\Delta+i0^+)\mathbb{I}-h_B(k))^{-1}_{\alpha_i \alpha_j} & = \frac{1}{\det(\Delta\mathbb{I}-h_B(k))}\nonumber\\
&\times \text{adj}(\Delta \mathbb{I}-h_B(k))_{\alpha_i \alpha_j}, \label{eq:inverse}
\end{align}
where $\text{adj}(\Delta\mathbb{I}-h_B(k))$ is the adjugate matrix, which is the transpose of cofactor matrix, which turns out to be built with the minors of $\Delta\mathbb{I}-h_B(k)$. Both the numerator and the denominator of \eqref{eq:inverse} are $\mathcal{O}(N_c)$-th degree polynomials of the matrix elements of $h_B(k)$, $\delta_\alpha(k)$ and $f_{\alpha\beta}(k)$ (Eq.~\eqref{eq:hBk}). As we are assuming that the hopping terms in the photonic bath are local up to a finite number of neighbours, then both $\delta_\alpha(k)$ and $f_{\alpha\beta}(k)$ can be written as finite sums of powers of $e^{\pm ik}$. This implies that the integrand of $J_{ij}$ is, up to a factor $e^{ikn_{ji}}$, a quotients of $\mathcal{O}(N_c)$-th degree polynomials of $e^{ik}$. Taking the change of variable $y=e^{ik\,\text{sgn}(n_{ji})}$, $J_{ij}$ transforms into:
\begin{equation}\label{eq:Jij_y}
J_{ij}=\frac{|g|^2}{2\pi i} \oint dy\; y^{|n_{ji}|} \frac{A_{\alpha_i\alpha_j}(y)}{B_{\alpha_i\alpha_j}(y)}.
\end{equation}

As a consequence of the previous discussion, both $A_{\alpha_i\alpha_j}(y)$ and $B_{\alpha_i\alpha_j}(y)$ are polynomials in $y$ with a finite degree $\mathcal{O}(N_c)$. As the poles of the integrand are inside or outside the unit circle, but never \emph{in} the circle (see App.~\ref{app:effective}), we can apply the residue's theorem taking into account just the poles inside the unit circle, $\{y_l^<\}$:
\begin{equation}\label{eq:Jij_sol}
J_{ij} = |g|^2 \sum_l  \text{Res}\left(\frac{A_{\alpha_i\alpha_j}(y) y^{|n_{ji}|}}{B_{\alpha_i\alpha_j}(y)},y=y_l^<\right).
\end{equation}
Since $B_{\alpha_i\alpha_j}(y)$ is a finite-order polynomial (see previous discussion) the degree of the poles $\{y_p^<\}$ is also finite. Using the definition of residue, each term in \eqref{eq:Jij_sol} will be proportional to $(y_p^<)^{|n_{ji}|}$. This means that irrespective of the model considered $J_{ij}$ \emph{is a finite sum of decaying exponentials} (notice that $|y_l^<|<1$), such that it can be written:
\begin{align}
J_{ij}= \sum_{l} C_l e^{-|n_{ji}|/\xi_l}\,\label{eq:sumaprox}
\end{align}
where the parameters $\{C_l,\xi_l\}$ depend on the particular model considered. Note, this provides a general no-go theorem for the possibility of having purely power-law interactions mediated by one-dimensional photonic models. In the next section, however, we will see how one can still mimic power-law interactions within certain distance regimes by breaking some of the assumptions of the generic model that we have considered.

\section{Emulating power-law interactions}\label{sect:power_law}

As mentioned in the introduction, long-range coherent interactions, understood as decaying with a power-law behaviour, can lead to qualitatively different phenomena in interacting spin models as compared to short-ranged ones. For example, these long-ranged models are known to break conventional Lieb-Robinson bounds on the spread of correlations~\cite{hauke13a,juneman13a,richerme14a,gong14a,fossfeig15a,gong16a}, to lead to exotic many-body phases~\cite{sandvik10a,islam13a,gong16a,gong16b,maghrebi17a,zunkovic18}, to modify the area-law~\cite{koffel12a,gong17a}, or to yield fast equilibration times~\cite{kastner11a}, among other phenomena. For this reason, it is still a relevant question to see whether in spite of the no-go theorem that we previously derived, it is still possible to obtain such power-law behaviour by breaking some of the assumptions of the generic model considered

In this section, we briefly sketch two possibilities: in~\ref{subsec:finexp}, we show how one can approximate power-law decay within a distance window by summing finite set of exponentials in a controlled fashion by going beyond local light-matter couplings. Then, in~\ref{subsec:long} we study baths with power-law hoppings, and show that they give rise to non-analytical energy dispersions, which allow one to obtain power-law asymptotic decays for the interactions (with the same decay exponent than the original hopping model).

\subsection{Mimicking power-laws adding finite number of exponentials ~\label{subsec:finexp}}

\begin{figure}[tbh!]
	\centering
	\includegraphics[width=0.95\columnwidth]{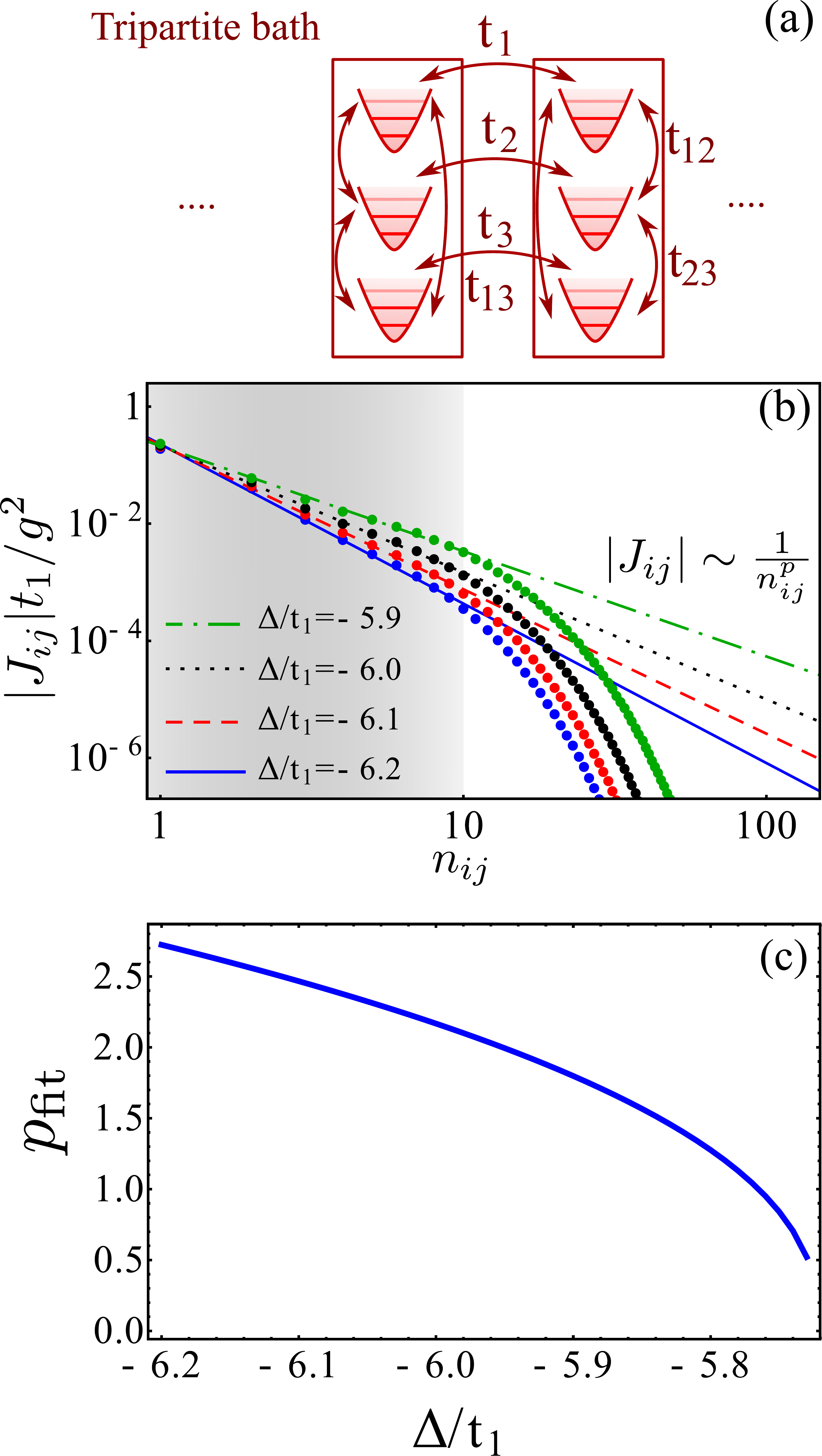}
	\caption{(a) General scheme of a tripartite lattice with hoppings defined in Hamiltonian of Eq.~\eqref{eq:hB3}. (b) Absolute value of $J_{ij}$ as a function of the emitter's distance $n_{ij}$ for two emitters coupled to sublattice $\alpha=1$. The points are the exact values and the lines the fits to a power law $\sim n_{ij}^{-p}$ for $1\leq n_{ij} \leq 10$. We take $t_1$ as the unit of energy. The couplings are $t_2=2.3 t_1$, $t_3=2.8t_1$, and $t_{12}=t_{23}=t_{13}=0.3t_1$. The values of $\Delta$ are in the legends. As shown there, $J_\text{eff}$ follows a power law up to $n_{ij}\simeq 10$ (shaded area). (c) Approximate power-law exponent as a function of $\Delta$ for the values of the parameters of panel (b). The exponent is obtained by a linear fit for small distances obtaining values $p_\mathrm{fit}\in(0.5,2.5)$.}
	\label{fig:tripartite}
\end{figure}

The fact that that power-laws can be built by summing up exponentials is well-known and used in many other fields (e.g., in economy~\cite{bochud06a}). In quantum optical setups, this has been already exploited, e.g., in Ref.~\cite{douglas15a}, to obtain approximate power-laws by the use of Raman-assisted transitions. As we show in the previous Section, in multipartite lattices such finite sums of exponentials appear in a natural way (see Eq.~\eqref{eq:sumaprox}), and can lead to power-law behaviour in certain distance windows. For example, taking a tripartite lattice described by a general Hamiltonian (see Fig.~\ref{fig:tripartite}(a)):
\begin{equation}\label{eq:hB3}
    h_B(k) = 
    \left(
    \begin{array}{ccc}
        -2t_1\cos k & t_{12} & t_{13}  \\
        t_{12} & -2t_2\cos k & t_{23}  \\
        t_{13} & t_{23} & -2t_3\cos k
    \end{array}
    \right),
\end{equation}
where $t_n$ is the nearest-neighbour hopping between the resonators in different sublattices, and $t_{\alpha\beta}$ describes the coupling between the $\alpha$-th and $\beta$-th resonator within the unit cells, we can find that $J_{ij}$ display a power-law behaviour in certain regimes. To illustrate it, we plot in Fig.~\ref{fig:tripartite}(b) the effective emitter interactions $J_{ij}$ for a given set of parameters (see caption), that we found mimic a power-law behaviour for distances up to $\sim 10$ neighbours. The different colors represent different values of the emitter's energy $\Delta$, which lead to different approximate power-law exponent. This is shown more explicitly in Fig.~\ref{fig:tripartite}(c) where we plot the results of the fitting $J_{ij}$ to a power-law for short distances. These interactions would already allow one to probe long-range interacting phenomena for small emitter's distances like done with trapped ions, e.g., in Refs.~\cite{islam13a,richerme14a}, where effective power-law behaviours were also obtained for chains of around 10 ions.  However, in order to fully explore the unconventional phenomena appearing in long-range interacting models, it would be desirable to extend its range to longer distances (ideally to the thermodynamic limit).

One way of extending the range of the power-law interactions would be to use the methods of optimal exponential expansions of power-law decays (see Ref.~\cite{bochud06a}). These methods provide a recipe to obtain the optimal $(C_l,\xi_l)$ parameters one needs to input in Eq.~\eqref{eq:sumaprox} to approximate a power-law decay with exponent $\nu$ up to a given distance, that are:
\begin{align}
C_{\alpha,\mathrm{opt}}&=\left(\frac{e}{\beta^\alpha}\right)^\nu\,,\label{eq:copt}\\
\xi_{\alpha,\mathrm{opt}}&=\frac{\beta^\alpha}{\nu}\,,\label{eq:xiop}
\end{align}
where $\beta$ is a number that must be optimized to match the power-law behaviour. Unfortunately, the relation between the physical parameters of generic multipartite lattices and the resulting parameters $(C_l,\xi_l)$ in Eq.~\ref{eq:sumaprox} is difficult to unravel, precluding their application.

\begin{figure}[tb]
	\centering
	\includegraphics[width=0.95\columnwidth]{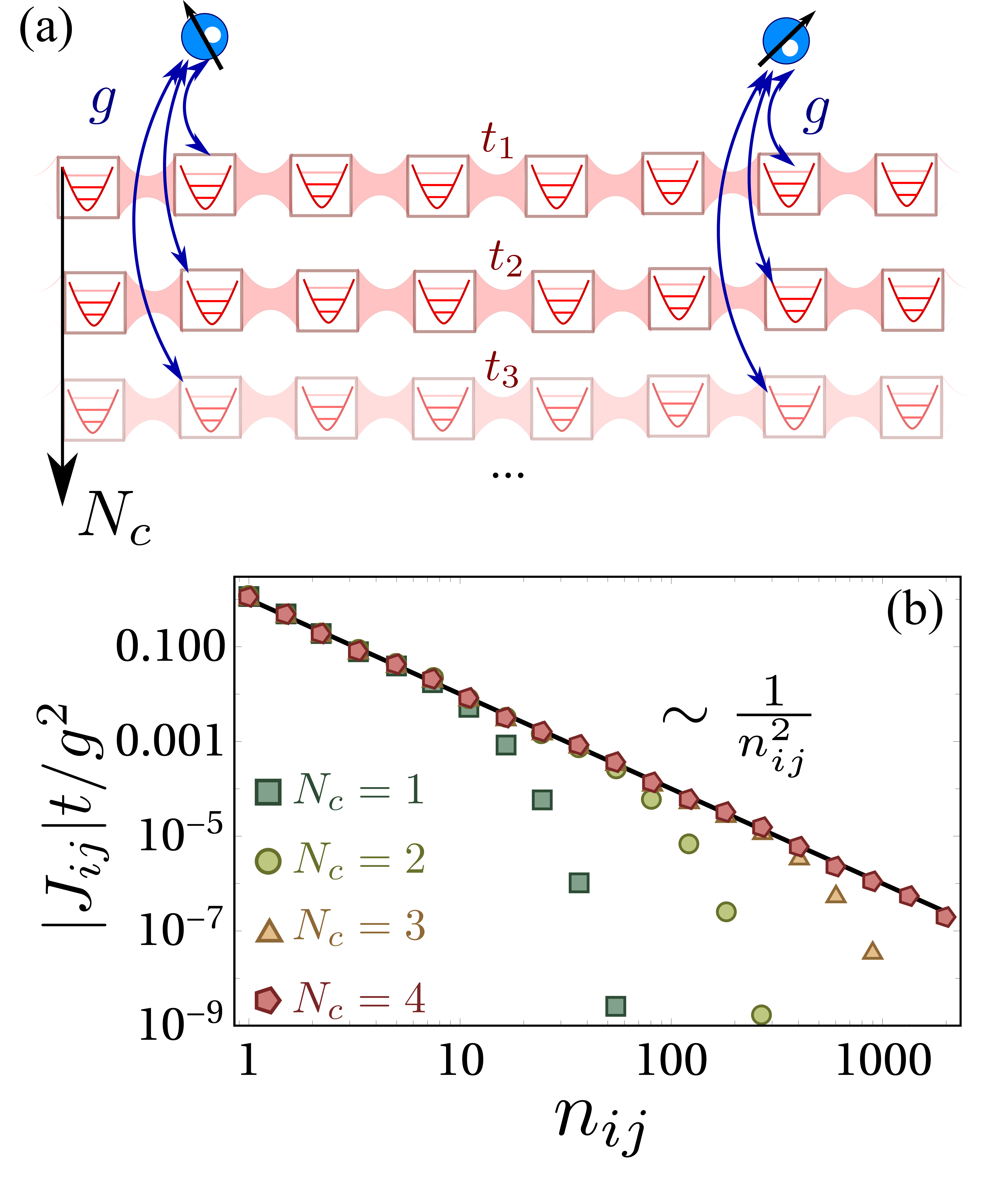}
	\caption{(a) Scheme of the non-local coupling of emitters to several uncoupled waveguides. (b) Result of applying the optimal expansion method of Ref.~\cite{bochud06a} to obtain $J_{ij}\approx 1/n_{ij}^2$ interactions. In solid black we plot the exact $1/n_{ij}^2$, while the markers are the results of approximation with $N_c=1, 2, 3, 4$ waveguides. The waveguide parameters have to be engineered such that the weigths and decay length match those of Eqs.~\eqref{eq:copt}-\eqref{eq:xiop}, using $\beta=6$.}
	\label{fig:nonlocal}
\end{figure}

 There is, however, one way where the application of these optimal expansion methods would be conceptually straightforward (although experimentally challenging). It requires allowing for non-local light-matter couplings, that is, that the emitters couple to several resonators $a_{n,\alpha=1,\dots,N_c}$ simultaneously. These couplings have been recently achieved with giant atoms in circuit QED platforms~\cite{kockum19a,kannan19a}. The idea consists in defining a multi-partite lattice with $N_c$ resonators per unit cell, but where they only interact with the resonators of different unit cells, that is, the bath consists of a set of uncoupled one-dimensional waveguides with energy dispersions $\omega_{\alpha}(k)=\tilde{\omega}_\alpha-2 t_\alpha \cos(k)$, where we write explicitly the possibility that the resonators of each uncoupled waveguide may have different energy $\omega_\alpha$. If the $j$-th emitter couples non-locally to all the resonators in a given unit cell $n_j$ (see Fig.~\ref{fig:nonlocal}(a)), the photon-mediated interactions will be given by the addition of the photon-mediated interactions of each uncoupled waveguide because $h_B(k)$ is diagonal (see App.~\ref{app:coupling}). Thus, $J_{ij}$ can be written as in Eq.~\eqref{eq:sumaprox}, but where $(C_{l},\xi_l)$ are given by (see Refs.~\cite{shi18a,calajo16a}):
\begin{align}
C_\alpha & \approx -\frac{g^2 \xi_\alpha}{2 t_\alpha}\,, \label{eq:nnmodel1}\\
\xi_\alpha & \approx \sqrt{\frac{t_\alpha}{D_\alpha}}\,,\label{eq:nnmodel2}
\end{align}
where $D_\alpha=\Delta-\tilde{\omega}_\alpha-2t_\alpha$ is the effective detuning of the emitter's energy to the lower-band-edge (we are assuming the emitter's frequency lie below the band-edge of all $\omega_\alpha(k)$). This means that by tuning $(t_\alpha,D_\alpha)$, one can tune independently the weight and range of the exponentials and match it to those of Eq.~\ref{eq:copt}-\ref{eq:xiop}. In Fig.~\ref{fig:nonlocal}, we illustrate this procedure by plotting how one can obtain a $J_{ij}\sim 1/n_{ij}^2$ interaction for increasing distance ranges up to $1000$ lattice sites by coupling the emitter non-locally to an increasing number of waveguides.

\subsection{Long-range hopping models~\label{subsec:long}}

\begin{figure}
	\centering
	\includegraphics[width=\linewidth]{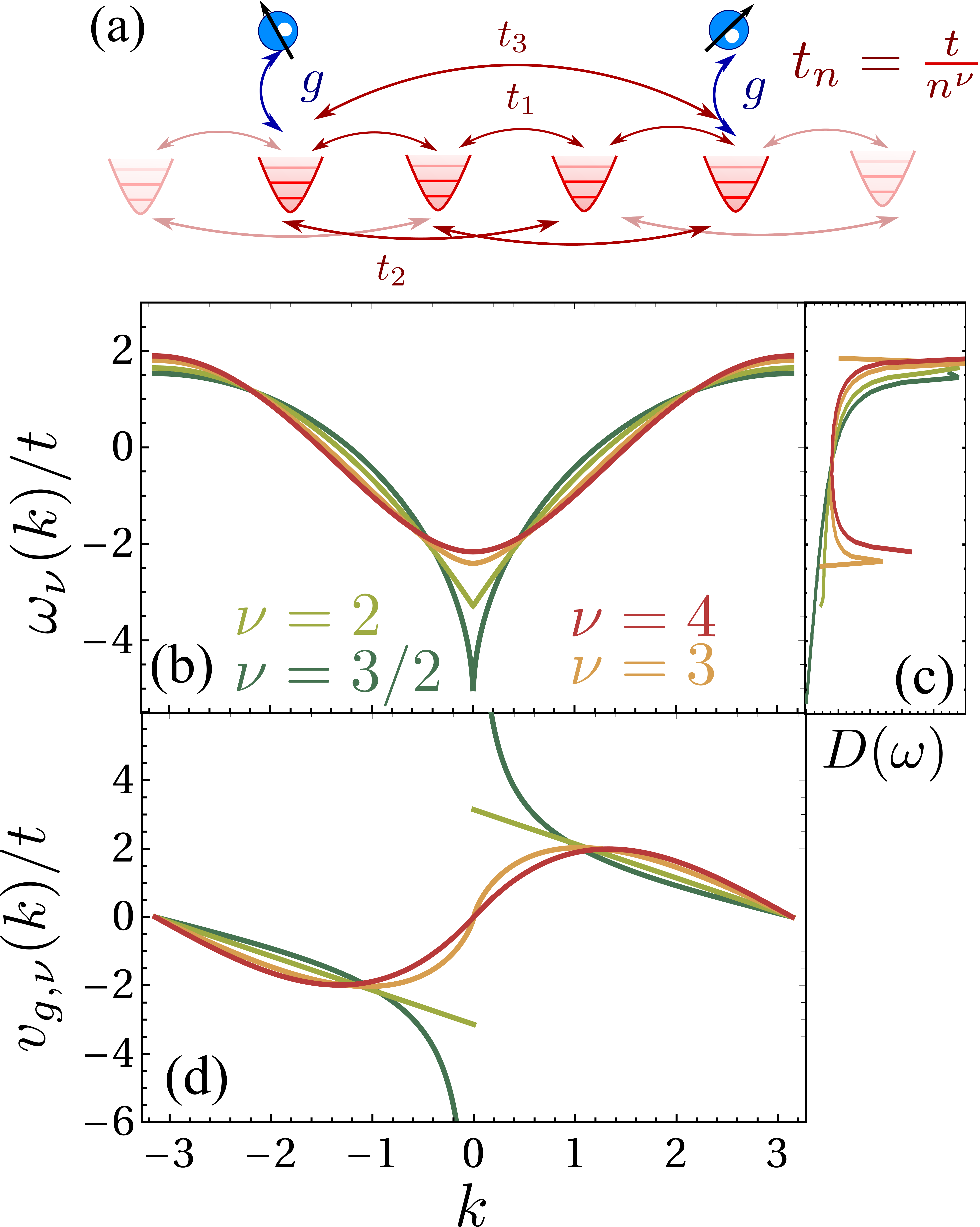}
	\caption{(a) Scheme of emitters coupled to a one-dimensional model with long-range hoppings $t_n=-\frac{t}{n^\nu}$ between $n$-neighbours. (b) Bath energy dispersion $\omega_\nu(k)/t$}
	\label{fig:dispersion_beta}
\end{figure}

The other possibility that goes beyond the initial bath assumptions of Sec.~\ref{sect:eff_int} consists in allowing for longer-range bath hoppings scaling as $t_n=t/n^\nu$, like depicted in Fig.~\ref{fig:dispersion_beta}(a). These long-range hopping models appear naturally in trapped ion systems~(see e.g., Ref~\cite{nevado16a}) and subwavelength atomic arrays~\cite{lehmberg70a,lehmberg70b}, showing both a power-law exponent of $\nu=3$.  They also appear in magnonic networks where superconducting loops can be used to enhance the range of the interactions to obtain $\nu<3$, as recently proposed in Ref.~\cite{rusconi19a}. In the spirit of the manuscript of providing results as general as possible we consider the situation where the exponent can have any value (limited by some physical bounds that we discuss afterwards).

The bath Hamiltonian of such photonic long-range hopping models can be described by a simple Bravais lattice with $N_c=1$, and its energy dispersion depends explicitly on the power-law exponent $\nu$:
\begin{equation}\label{eq:dispersion_beta}
    \omega_\nu(k) = -t(\text{Li}_\nu(e^{ik})+\text{Li}_\nu(e^{-ik})).
\end{equation}

Here, $\text{Li}_\nu(z)$ is the polylogarithm function of order $\nu$. These energy dispersions are finite for every $k$ as long as $\nu>1$, since $\omega_1(k)$ displays a logarithmic singularity around $k=0$. Thus, we will restrict our discussion to models with $\nu>1$. As an illustration, in Figs.~\ref{fig:dispersion_beta}(b-c) we plot the energy dispersion and associated density of states $D(\omega)$ for models with power-law exponents $\nu=3/2,2,3$ and $4$. As expected for large exponents, both the energy dispersion and density of states tend to converge to the nearest-neighbour case of $\omega(k)\approx -2t\cos(k)$, with two van-Hove singularities at the band-edges~(see, e.g.,~\cite{calajo16a,shi16a,shi18a}). When the exponent decreases, however, the longer-range hoppings strongly modify the band structure and associated density states. In particular, we observe that $\omega_\nu(k)$ features a visible non-analytical kink around $k\approx 0$, which is more evident when we calculate the group velocity of the model, i.e., $v_{g,\nu}(k)=\partial_k \omega_{\mu}(k)$, which we plot in Fig.~\ref{fig:dispersion_beta}(d) for the same power-law exponents. There, we observe, for example, the finite discontinuous jump of the group velocity for $\nu=2$, i.e.,  $v_{g,2}(k)\approx \pi \mathrm{sign}(k)-k$. The jump becomes bigger as $\nu\rightarrow 1$, finally showing a $1/k$ divergence when $\nu=1$. This increase of the group velocity around $k=0$ leads to strong modification of the density of states, e.g., canceling the lower edge Van-Hove singularity as $\nu\rightarrow 1$.  

\begin{figure}
    \centering
    \includegraphics[width=\linewidth]{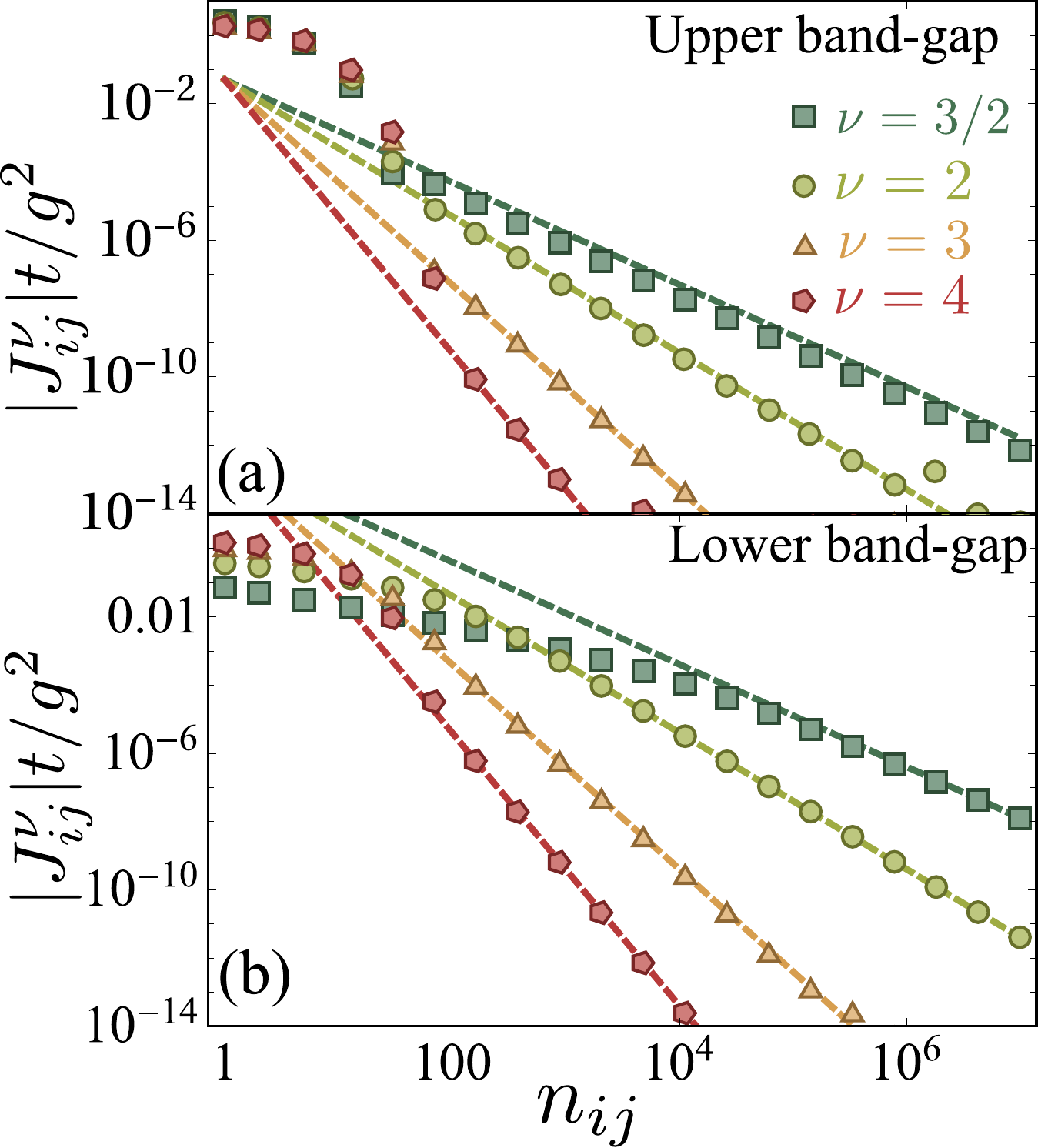}
    \caption{Effective emitter's interaction $J^\nu_{ij}$ induced by the photonic long-range hopping models as a function of the emitter's distance $n_{ij}$ for $\nu=1.5$, $2$, $3$, and $4$ (see legend). The markers are the result of the numerical integration of Eq.~\eqref{eq:Jijlong}, whereas the dashed lines are the asymptotic scaling laws that match those of the original hopping model. We fix the detuning between the emitter's energy and the bottom/top of the band to $0.05J$: $\Delta=\omega_\nu(0/\pi)\pm 0.05J$ in panels (a) and (b), respectively.}
    \label{fig:alpha_beta}
\end{figure}

Like in other structured baths~\cite{john94a}, such non-analytical behaviour of the density of states will result in non-Markovian quantum dynamics when the emitter's frequencies are tuned with the non-analytical regions. In this manuscript, however, we will focus only on characterizing the effective emitter's interactions $J_{ij}$ in the regime where one can still adiabatically eliminate the photonic bath (Born-Markov regime) by assuming $\Delta$ lies far enough from the band edges~\cite{calajo16a,shi16a,shi18a}. Since this bath can be written as a simple Bravais lattice, the expression of Eq.~\eqref{eq:Jij} simplifies to:
\begin{equation}\label{eq:Jijlong}
J^\nu_{ij} = \frac{|g|^2}{2\pi}\int_{-\pi}^\pi dk \frac{e^{ik n_{ji}}} {\Delta-\omega_\nu(k)}\,.
\end{equation}

Differently from the baths we have considered up to now, the $\omega_\nu(k)$ energy dispersion has a branch-cut along the imaginary axis that will lead to qualitatively different behaviour (see App.~\ref{app:long} for a complete discussion of the integration).  To characterize $J_{ij}^\nu$ for all distance regimes, it is convenient to distinguish the situations when the emitter's energy lies in the upper/lower band-gap region. 

$\bullet$ \emph{Upper band-gap regime.} This corresponds to situations when $\Delta>\omega_\nu(\pi)$. In that case, the photon-mediated interactions can always be written as a sum of two contributions:
\begin{equation}
J_{ij}^\nu=J_{ij,\mathrm{pole}}^\nu+J^\nu_{ij,\mathrm{BC}}\,,
\end{equation}
where $J_{\mathrm{pole}}^\nu$ is the contribution of the poles of the denominator of the integrand of Eq.~\ref{eq:Jijlong}. Their contribution can be obtained using Residue theorem by expanding $\omega_\nu(k)$ close to the position where the pole is expected, i.e., $\omega_\nu(\pm \pi +i y)/t\approx \omega_\nu(\pi)/t+A_\nu y^2$, for $y\ll 1$. Using that expansion, we find that:
\begin{align}
J_{ij,\mathrm{pole}}^\nu\approx (-1)^{|n_{ij}|}\frac{|g|^2  \xi_{\nu}}{2t A_\nu} e^{-|n_{ij}|/\xi_\nu}\,,
\end{align}
where $\xi_\nu=\sqrt{A_\nu t/D_u}$, $D_u=\Delta-\omega_\nu(\pi)$, and $A_\nu$ is a numerical constant that depends on the exponent $\nu$. The other contribution to $J_{ij}^\nu$ comes from the detour we have to take when using Residue theorem to avoid the branch-cut of the integrand. This contribution can be written as:
\begin{align}
J_{ij,\mathrm{BC}}^\nu  &\approx- \frac{|g|^2}{\pi}\int_0^\infty dy \frac{ \mathrm{Im}\left[\omega_\nu(\varepsilon+i y)\right]e^{-y|n_{ij}|}}{\left(\Delta-\mathrm{Re}\left[\omega_\nu(\varepsilon+i y)\right]\right)^2+\mathrm{Im}\left[\omega_\nu(\varepsilon+i y)\right]^2}=\label{eq:JijBCl}\\
&=\int_0^\infty G_\nu(y) e^{-y|n_{ij}|}\,.
\end{align}

Irrespective of $\Delta$, this term will eventually dominate the long-distance behaviour of $J_{ij}^\nu$ since the pole contribution is exponentially attenuated. Due to the exponential term $e^{-y|n_{ij}|}$ of the integrand of Eq.~\ref{eq:JijBCl}, the long distance behaviour of $J_{ij,\mathrm{BC}}^\nu$ will be dominated by the behaviour of $G_\nu(y)$ when $y\ll 1$, which we find to be: $G_\nu(y\ll 1)\propto y^{\nu-1}$. Using that  $\int_0^\infty y^\alpha e^{-y d}=\Gamma(1+\alpha)/d^{1+\alpha}$, we can show then that the photon-mediated interactions scale as:
\begin{align}
J_{ij,\mathrm{BC}}^\nu \propto -\frac{|g|^2 t}{D_u^2}\frac{1}{|n_{ij}|^\nu}\,,
\end{align}
in the asymptotic limit $(|n_{ij}|/\xi_\nu\gg 1)$. Thus, the dipole-dipole interactions $J_{ij}^\nu$ inherit the power-law exponent from the hopping model. This behaviour is illustrated in Fig.~\ref{fig:alpha_beta}(a) where we plot the result of numerically integrating $J^\nu_{ij}$ as a function of the emitter's distance $n_{ij}$ for the same power-law exponents chosen for Fig.~\ref{fig:dispersion_beta}, and for a detuning $D_u=0.05t$. There, we clearly observe that after an initial exponential decay of the interactions coming from the pole contribution, $J_{ij}^\nu$ features an asymptotic power-law scaling (dashed line) with the same power-law exponent than the original bath model. Let us note, that this regime was already explored for $\nu=3$ in the context of trapped ions~\cite{nevado16a} obtaining similar results. 

$\bullet$ \emph{Lower band-gap regime.} This regime corresponds to situations when $\Delta<\omega_\nu(0)$, and remarkably it leads phenomena that, to our knowledge, has not been pointed out before. The main difference with respect to the upper-bandgap situation is that the pole contribution can be shown to be strictly zero $J^\nu_{\mathrm{pole}}\equiv 0$, such that $J_{ij}^\nu$ comes solely branch-cut detour contribution $J_{ij,\mathrm{BC}}^\nu$ (see Appendix). This makes its short-distance behaviour strongly dependent on the particular $\nu$-exponent of the hopping model and qualitatively very different from the upper band-gap situation. This is clearly seen in Fig.~\ref{fig:alpha_beta}(b), where we plot the $J_{ij}^\nu$ for the same detunings and exponent than in panel (a), but for the lower band-gap. There, we observe that:
\begin{itemize}
	\item For $\nu \leq 3$ the $J_{ij}^\nu$ does not display the initial exponential decay coming from the pole contribution. In fact, for the case of $\nu=2$, we can find that the initial decay follows a logarithmic law:
	\begin{align}
	J_{ij}^{\nu=2}\approx \frac{|g|^2\left[\gamma+\log(|n_{ij}|D_l/(\pi t))\right]}{\pi^2 t}\,,~\label{eq:Jij22main}
	\end{align}
	for distances $|n_{ij}|\ll \pi t/D_l$. For $\nu=3$ we also derived a (more cumbersome) analytical expression (see Appendix) which shows a similar logarithmic decay.
	
	\item For $\nu>3$, even though $J_{ij}^\nu$ comes entirely from the branch-cut contribution of Eq.~\ref{eq:JijBCl}, it starts displaying an approximated exponential decay (see Fig.~\ref{fig:alpha_beta}(b)).  The mathematical reason is that one can find an approximated pole of Eq.~\ref{eq:Jijlong}, whose residue can be approximated by:
\begin{align}
\frac{J_\mathrm{ij,pole,*}^\nu t}{|g|^2}\approx \frac{\xi^*_\nu}{2 C_\nu} e^{-|n_{ij}|/\xi^*_\nu}\label{eq:Jijpole2main}
\end{align}
with $\xi_{\nu}^*=\sqrt{|C_\nu| t/D_l}$ with $D_l=\omega_{\nu}(0)-\Delta>0$, and $C_\nu$ coming from the expansion $\omega_\nu(0^++iy)/t\approx\omega_\nu(0)/t+C_\nu y^2$ when $y\ll 1$. When $\nu\gg 1$, $C_\nu\approx -1$, recovering the results of the nearest-neighbour model that were given in Eqs.~\ref{eq:nnmodel1}-\ref{eq:nnmodel2}. Note, that this is expected since when $\nu\gg 1$ the longer range hoppings are negligible with respect to the nearest neighbour ones.
	
\end{itemize}

\section{Conclusions}\label{sect:conclusions}

Summing up, we have derived several general results for the limits of (coherent) photon-mediated interactions induced by one-dimensional photonic environments. First, we have shown that under the standard assumptions of locality of light-matter (rotating-wave) couplings and photon hoppings, the photon-mediated interactions can always be written as a finite sum of exponentials. Thus, they can only display power-law behaviour in small distance windows. Besides, we have also proposed two ways of extending the range of such power-law behaviour by considering models that go beyond the previous assumptions. For example, by coupling non-locally to several one-dimensional waveguides, one could extend considerably the range of the power-law behaviour of the interactions in a controlled fashion. Finally, we have also considered the photon-mediated interactions appearing in one-dimensional baths with power-law hoppings. These models display non-analytical energy dispersions which lead to interactions which inherit the asymptotic scaling of the original hopping model. Besides, we also find situations where the photon-mediated interactions are not exponentially attenuated in any distance window. We foresee that other baths with similar non-analytical energy dispersions, such as resonators arrays coupled with $X_i X_j$ couplings or critical spin baths, will also display similar power-law asymptotic scalings. Another interesting direction to explore is whether these conclusions hold as well for models in the ultra-strong coupling regime, where such photon-mediated interactions have recently started being explored~\cite{roman20a}.

\section{Acknowledgments}

We acknowledge Ignacio Cirac, Johannes Knörzer, Daniel Malz, Martí Perarnau, and Cosimo C. Rusconi for inspiring and fruitful discussions. Eduardo Sánchez-Burillo acknowledges ERC Advanced Grant QUENOCOBA under the EU Horizon 2020 program (grant agreement 742102). AGT acknowledges funding from project PGC2018-094792-B-I00  (MCIU/AEI/FEDER, UE), CSIC Research Platform PTI-001, and CAM/FEDER Project No.~S2018/TCS-4342~(QUITEMAD-CM).

\newpage
\appendix

\section{Introduction}

In this Appendix we provide more details on: i) the general diagonal form of the bath Hamiltonian in section~\ref{app:HB}; ii) the derivation of the general photon-mediated interactions $J_{ij}$ in section~\ref{app:effective}; iii) the generalization of $J_{ij}$ to the case in which the light-matter couplings are not fully local, in section~\ref{app:coupling}; and  finally, iv) the detailed analysis of the photon-mediated interactions for the long-range hopping model in section~\ref{app:long}.

\section{Diagonal form of $H_B$}\label{app:HB}

The bosonic Hamiltonian $H_B$ of Eq. \eqref{eq:H} can be diagonalized. As $h_B(k)$ is Hermitian:
\begin{equation}\label{eq:PDP}
h_B(k) = P(k) D(k) P^\dagger (k),
\end{equation}
being $P(k)$ unitary and 
\begin{equation}
D(k)=\text{diag}(\omega_1(k),\omega_2(k),\dots,\omega_{N_c}(k)),
\end{equation}
with $\{\omega_n(k)\}_{n=1}^{N_c}$ being the eigenvalues of $h_B(k)$. Then, defining a new set of bosonic operators $\{\hat{\alpha}_{n,k}\}_{n=1}^{N_c}$
\begin{equation}\label{eq:alpha}
    \left(
    \begin{array}{c}
         \hat{\alpha}_{1,k} \\
         \vdots \\
         \hat{\alpha}_{N_c,k}
    \end{array}
    \right) \equiv 
    P^\dagger(k)
    \left(
    \begin{array}{c}
         \hat{a}_{1,k} \\
         \vdots \\
         \hat{a}_{N_c,k}
    \end{array}
    \right).
\end{equation}

$H_B$ then reads
\begin{equation}
    H_B = \sum_k \sum_{n=1}^{N_c} \omega_n(k) \hat{\alpha}_{n,k}^\dagger  \hat{\alpha}_{n,k}.
\end{equation}

\section{Effective photon-mediated interactions}\label{app:effective}

In order to derive Eq.~\eqref{eq:Jij} from Eq.~\eqref{eq:Jij_E}, we write the bosonic operators $a_{n_j,\alpha}$ of $H_\text{int}$ in momentum space and in terms of the $\alpha$-modes of Eq.~\eqref{eq:alpha}:
\begin{align}
    a_{n_j,\alpha}&=\frac{1}{\sqrt{N}}\sum_k e^{ikn_j}\hat{a}_{k,\alpha} \nonumber \\
    &= \frac{1}{\sqrt{N}}\sum_k e^{ikn_j}\sum_{\beta=1}^{N_c} P_{\alpha_j \beta}^*(k) \hat{\alpha}_{k,\beta}.
\end{align}

From this and Eq.~\eqref{eq:Hint}, the only eigenvalues contributing to the sum in \eqref{eq:Jij_E} are $\ket{E}=\hat{\alpha}_{k,n}^\dagger\ket{0}$, with $E=\omega_n(k)$. After taking the thermodynamic limit $N\to\infty$ in Eq.~\eqref{eq:Jij_E}:
\begin{equation}\label{eq:Jij_P}
    J_{ij} = \frac{|g|^2}{2\pi}\int_{-\pi}^\pi dk \sum_{\beta=1}^{N_c} \frac{P_{\alpha_i \beta}(k) P^*_{\alpha_j\beta}(k)}{\Delta+i0^+-\omega_\beta(k)} e^{ikn_{ji}}.
\end{equation}
The sum in the integrand can be rewritten as:
\begin{equation}\label{eq:sum}
    \sum_{\beta,\gamma=1}^{N_c} P_{\alpha_i \beta}(k) \frac{\delta_{\beta\gamma}}{\Delta+i0^+-\omega_\beta(k)}(P^\dagger(k))_{\gamma \alpha_j}.
\end{equation}

Notice that $P(k)$ is the matrix which diagonalizes $h_B(k)$ (Eq.~\eqref{eq:PDP}). Using the well-known property
\begin{align}
& P(k) D(k)P^\dagger(k) = h_B(k) \nonumber\\
\Rightarrow & P(k) f(D(k)) P^\dagger(k) = f(h_B(k))
\end{align}
we see that \eqref{eq:sum} is $((\Delta+i0^+)\mathbb{I}-h_B(k))^{-1}_{\alpha_i\alpha_j}$. Introducing this in \eqref{eq:Jij_P}, we get \eqref{eq:Jij}.

Besides, notice that the inverse $((\Delta+i0^+)\mathbb{I}-h_B(k))^{-1}$ always exists. To prove that, we write down the determinant of $(\Delta+i0^+)\mathbb{I}-h_B(k)$:
\begin{equation}
    \det((\Delta+i0^+)\mathbb{I}-h_B(k)) =  \prod_{\beta=1}^{N_c} (\Delta+i0^+ - \omega_\beta(k)).
\end{equation}
As $\Delta$ is not embedded in the bands: $\Delta\neq \omega_\alpha(k)$ $\forall k,\alpha$, then this determinant is different from 0. In consequence, its inverse exists.

A corollary of this last result is that the integrand of $J_{ij}$ (Eq.~\eqref{eq:Jij_y}) has no poles in the unit circle, $|y_l|\neq 1$. This is due to the fact that the polynomial of the denominator of the integrand of $J_{ij}$ is proportional to the determinant of $(\Delta+i0^+-h_B(k))^{-1}$ (see Eqs.~\eqref{eq:Jij},~\eqref{eq:inverse}, and~\eqref{eq:Jij_y}). We just proved that,  provided $\Delta$ is not in the bands of the model, this determinant is different from 0 $\forall k\in \mathbb{R}$, that is, for $|y|=1$. In consequence, all the poles are inside or outside the circle, but never \emph{in} the circle.

\section{Beyond point-like light-matter couplings}\label{app:coupling}

Let us assume that each emitter couples to a finite number of contiguous sites:
\begin{equation}
     H_\text{int} = \sum_{j=1}^2 \sigma_j^+ \sum_{\alpha,n} g^j_{\alpha,n} a_{n,\alpha} + \text{H.c.}
\end{equation}

Introducing this interaction Hamiltonian in the definition of $J_{ij}$,~\eqref{eq:Jij_E}, and taking the thermodynamic limit
\begin{align}
\label{eq:Jij_nl}
J_{ij} & =\frac{1}{2\pi}\sum_{\alpha_i\alpha_jn_in_j}(g^i_{\alpha_in_i})^* g^j_{\alpha_jn_j} \nonumber \\
&\times \int dk\, e^{ik(n_j-n_i)}((\Delta+i0^+)\mathbb{I}-h_B(k))_{\alpha_i\alpha_j}^{-1}.
\end{align}
The integrand is identical to the point-like case considered in the main text (Eq. \eqref{eq:Jij}), being the only difference the sum over the couplings to the different sites and sublattices. Therefore, $J_{ij}$ is again a sum of exponentials of $|n_{ji}|$, weigthed each with the couplings $(g^i_{\alpha_in_i})^* g^j_{\alpha_jn_j}$.

As a corollary of this, we can derive the effective interaction in the particular case in which there is no interaction in each sublattice, so $f_{\alpha\beta}(k)=0$ and $\omega_\alpha(k)=\delta_\alpha(k)$ (see Eq.~\eqref{eq:hBk}), and the $i$-th qubit is coupled to the position $n_i$ of each sublattice with the same coupling strength $g$. Considering this in Eq.~\eqref{eq:Jij_nl}, we get
\begin{equation}
  J_{ij} = \frac{|g|^2}{2\pi}\sum_\alpha \int_{-\pi}^\pi dk \frac{e^{in_{ji}}}{\Delta-\omega_\alpha(k)}\,,
\end{equation}
that is, that the effective photon-mediated interactions is the sum of the ones induced independently by each energy band $\omega_\alpha(k)$

\section{Asymptotic scaling of the interactions in long-range photonic models}\label{app:long}

As we have shown in the main text, the energy dispersion of the photonic bath model with power-law hoppings of exponent $\nu$ is given by:
\begin{align}
\omega_\nu(k)=-t\left(\mathrm{Li}_\nu(e^{ik})+\mathrm{Li}_\nu(e^{-ik})\right)\,,
\end{align}
where $\mathrm{Li}(y)$ is the polylogarithm function. This function is defined by a power-series:
\begin{align}
\mathrm{Li}_\nu(z)=\sum_{k=1}^\infty \frac{z^k}{k^\nu}\,,
\end{align}
valid for any complex $\nu$ and all complex arguments within $|z|<1$, although it can be analytically continued to the whole complex plane. Assuming a local coupling to the bath, with coupling strength $g$, the photon mediated interactions are given by:
\begin{align}
\frac{J^\nu_{ij}}{|g|^2}=\frac{1}{2\pi}\int_{-\pi}^{\pi}dk\frac{e^{i k|n_{ij}|}}{\Delta-\omega_\nu(k)}\,,\label{eq:jijmu}
\end{align}

In the case of case of finite-range hopping models, this integral was calculated by making the change of variable $z=e^{ik}$ and was shown to be given solely the contribution of the poles within the unit circle. Here instead, the contour can not be simply closed because $\omega_\nu(k)$ has a branch-cut along the imaginary axis of $\mathrm{Im}k\in (-\infty,\infty)$ inherited by the branch-cut of the polylogarithm function. As a consequence, $J^\nu_{ij}$ will contain additional contributions coming from the detours to avoid the branch-cut. Let us now provide approximated expressions of $J_{ij}^\nu$ for both the situation where the emitters' energy lie above and below the band, that we will show to lead to qualitatively different behaviour.

\emph{Upper band-gap.} Let us start with the case where $\Delta>\omega_{\nu}(k)$ for all $k$, and denote as $D_u$ the energy difference between the emitter's energy and the upper band-edge, i.e, $D_u=\Delta-\omega_{\nu}(\pm \pi)>0$. Taking $k$ as a complex variable, we can find that the denominator of the integrand of Eq.~\ref{eq:jijmu} has four complex poles close to $\pm \pi$ but slightly above/below the real axis. To find an approximated expression for the poles, we can expand $\omega_{\nu}(k)$ close to that point, finding it can always be written as:
\begin{align}
\omega_{\nu}(\pm\pi +i y)/t\approx \omega_\nu(\pm \pi)/t +A_\nu y^2\,,
\end{align}
for $|y|\ll 1$, and where the higher-order terms of the expansion remain real. This means that to the lowest order of $y$, the poles can be approximated by: 
\begin{align}
k_{\pi,\pm} \approx \pi\pm i\sqrt{D_u/A_\nu}\,,\\
k_{-\pi,\pm} \approx -\pi\pm i\sqrt{D_u/A_\nu}\,.
\end{align}

\begin{figure}
	\centering
	\includegraphics[width=\linewidth]{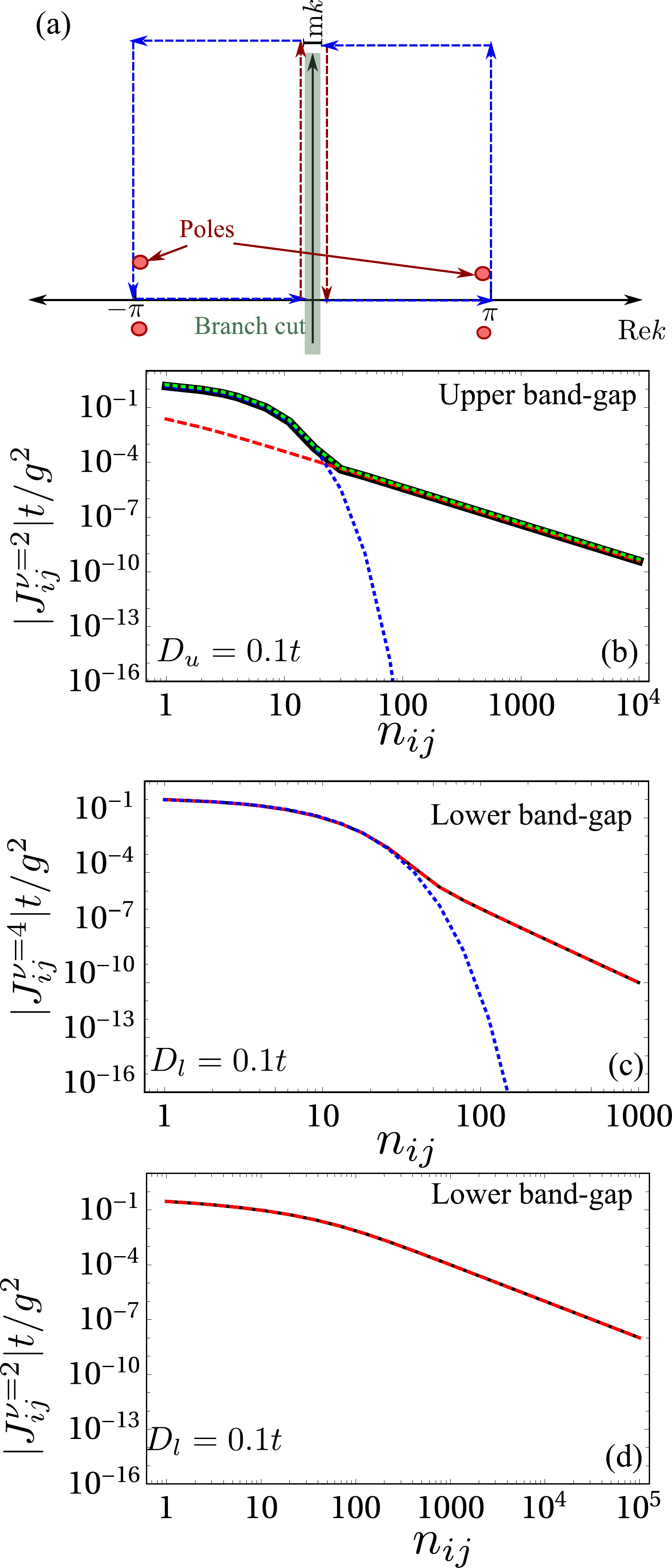}
	\caption{(a) Contour of integration (in dashed blue) to obtain $J_{ij}^\nu$ an approximated expression of Eq.~\eqref{eq:jijmu}. (b) $J_{ij}^{\nu=2}$ as a function of the distance $n_{ij}$ for an emitter tuned to the upper band-gap. We also plot separately the contribution of the poles (dotted blue) and the branch-cut detour one (dashed red). In dotted green we plot the sum of the two contributions given by Eqs.~\eqref{eq:Jijpole} and~\eqref{eq:JijBC}, respectively. (c)  $J_{ij}^{\nu=4}$ calculated from numerically integrating Eq.~\eqref{eq:jijmu} for an emitter tuned to the lower-band gap. In dotted blue, we plot the approximated expression found in Eq.~\eqref{eq:Jijpole2}. (d) $J_{ij}^{\nu=2}$ calculated from numerically integrating Eq.~\eqref{eq:jijmu} for an emitter tuned to the lower-band gap. In dashed red, we plot the analytical expression obtained in Eq.~\eqref{eq:Jij2}. }
	\label{figap:contour}
\end{figure}

Thus, if we define the contour as depicted with dashed arrows in Fig.~\ref{figap:contour}(a), the integral of Eq.~\eqref{eq:jijmu} can be shown to be given by two contributions: the one of the poles embedded within the contour, plus the ones of the two detours along the branch-cut (red arrows):
\begin{equation}
J_{ij}^\nu= J_{ij,\mathrm{pole}}^\nu+J_{ij,\mathrm{BC}}^\nu\,.
\end{equation}

The only poles that contribute are the ones with positive imaginary part, i.e., $k_{\pm \pi,+}$, whose contribution can be obtained by calculating their associated residue. This can be done by writing $\Delta-\omega_\nu (k)\approx A_\nu(k-k_{\pm \pi,+})(k-k_{\pm\pi,-})$ for the $k$'s close to the poles, which yields
\begin{equation}
\frac{J_{ij,\mathrm{pole}}^\nu t}{g^2}\approx \frac{(-1)^{|n_{ij}|}}{2}\frac{\xi_\nu}{A_\nu} e^{-|n_{ij}|/\xi_\nu}\label{eq:Jijpole}\,.
\end{equation}
where $\xi_u=\sqrt{A_\nu t/D_u}$ provides the localization length of the interaction and its strength. Note that because the poles lie in the along the integration contour, they contribute with half of its residue. This can be proven more rigorously by making the integration along this line and making use of the identity $\frac{1}{x\pm i0^+}=\mp i\pi \delta(x)+\mathrm{P}\left(1/x\right)$, or by shifting the domain of integration of Eq.~\eqref{eq:jijmu} to $(0,2\pi)$ as done in Ref.~\cite{nevado16a}.

 The branch-cut detour contribution can be written in the following compact expression:
\begin{align}
\frac{J_{\mathrm{BC}}^\nu t}{g^2}&\approx -\frac{1}{\pi}\int_0^\infty dy \frac{ \mathrm{Im}\left[\omega_\nu(\varepsilon+i y)\right]e^{-y|n_{ij}|}}{\left(\Delta-\mathrm{Re}\left[\omega_\nu(\varepsilon+i y)\right]\right)^2+\mathrm{Im}\left[\omega_\nu(\varepsilon+i y)\right]^2}=\label{eq:JijBC}\\
&=\int_0^\infty G_\nu(y) e^{-y|n_{ij}|}\,.
\end{align}

Since the pole contribution is exponentially damped for larger distances, the asymptotic scaling of $J_{ij}^\nu$ will always be provided by $J_{\mathrm{BC}}^\nu$. Due to the $e^{-y|n_{ij}|}$ dependence of the integrand, the $J_{\mathrm{BC}}^\nu$ behaviour at long distances is dominated by the dependence of $G_\nu(y)$ for $y\ll 1$, which we find to be: $G_{\nu}(y)\approx B_\nu y^{\nu-1}$, with $B_\nu$ being a constant that depends on $\Delta$ and the exponent $\nu$. Since $\int_0^\infty y^\alpha e^{-y d}=\Gamma(1+\alpha)/d^{1+\alpha}$ for $\alpha>-1$, $d>0$, and $\Gamma(x)$ being the $\Gamma$-function, the final asymptotic scaling of the photon-mediated interactions  can be shown to inherit the same power-law behaviour of the hopping model, i.e., $J_{ij}^\nu\propto 1/|n_{ij}|^\nu$ for $n_{ij}/\xi_{\nu}\gg 1$.

As an illustration that the expressions derived above reproduce well the behaviour of $J_{ij}^\nu$, in Figs.~\ref{figap:contour}(b) we plot together the $J_{ij}^\nu$ for $\nu=2$ obtained from the direct numerical integration of Eq.~\ref{eq:jijmu} (in solid black), and its different contributions: in dotted blue the pole contribution given by Eq.~\ref{eq:Jijpole}, in dashed red the branch-cut contribution as defined in Eq.~\ref{eq:JijBC}, and in dotted green the sum of the two. The other power-law exponents lead to qualitatively similar phenomena.

\emph{Lower band-gap.} When the emitter's energy lies in the lower band-gap, that is, $\Delta<\omega_\nu(k)$ the denominator has no pole, such that $J_{ij}^\nu$ is only given by the branch-cut contribution, i.e., $J^\nu=J_{\mathrm{BC}}^\nu$. The underlying reason in these region one should try to find the poles close to zero momentum, e.g., $k=0^{\pm}+iy$. However, when expanding $\omega_\nu(k)$ around it we find that differently from $\omega_\nu(\pm \pi + iy)$, it contains both real and imaginary terms, such that no solution can be found. In particular, we find that (for integer $\nu$):
\begin{align}
\mathrm{Re}\omega_\nu(0^+ +iy)&\approx \omega_\nu(0)+C_\nu y^2\,,\label{eq:aux1}\\
\mathrm{Im}\omega_\nu(0^+ +i y)&\approx D_\nu y^{\nu-1}\label{eq:aux2}\,.
\end{align}
with $D_\nu>0$ for all $\nu$, but where $C_\nu$ behaves differently depending on $\nu$:
\begin{align}
C_{2}&=\frac{1}{2}\,, \\
C_{3}&=\left(\ln(y)-\frac{3}{2}\right)\,,\label{eq:c3}\\
C_{\nu>3}&<0\,.\label{eq:aux3}
\end{align}

The different behaviour of such expansions for the different exponents will lead to qualitatively different photon-mediated interactions $J_{ij}^\nu$. Thus, it is convenient to analyze separately the different situations that can appear:

$\bullet$ When $\nu>3$, the $J^\nu_{ij}$ for short distances starts to features an exponential decay like in the upper band-gap situation. This is illustrated in Fig.~\eqref{figap:contour}(c) for $\nu=4$.  Note, that this behaviour is expected because when $\nu\gg 1$, one should recover the limit of nearest-neighbour hoppings which feature solely an exponential decay (Eqs.~\eqref{eq:nnmodel1}-\eqref{eq:nnmodel2}). This can be reconciled mathematically by noticing that when $\nu>3$, the imaginary contribution of $\omega_\nu(0^++iy)$ starts to be subleading compared to the real part (see Eqs.~\eqref{eq:aux1}-\eqref{eq:aux2}). In that case, if one neglects this imaginary part, the integrand of $J_{ij}^\nu$ in Eq.~\eqref{eq:jijmu} will have a pole, whose contribution can be approximated by:
\begin{align}
\frac{J_{ij,\mathrm{pole,*}}^\nu t}{|g|^2}\approx \frac{\xi^*_\nu}{2 C_\nu} e^{-|n_{ij}|/\xi^*_\nu}\label{eq:Jijpole2}
\end{align}
with $\xi_{\nu}^*=\sqrt{-C_\nu t/D_l}$ with $D_l=\omega_{\nu}(0)-\Delta>0$ (remember that $C_{\nu>3}<0$, see Eq.~\ref{eq:aux3}). For $\nu\gg 1$, $C_\nu\approx -1$, recovering the results of the nearest-neighbour model that were given in Eqs.~\eqref{eq:nnmodel1}-\eqref{eq:nnmodel2}.

$\bullet$ When $\nu<3$, on the contrary, the imaginary part of $\omega_\nu(0^+ +i y)$ is of higher order in $y$ than the real one (see Eqs.~\eqref{eq:aux1}-\eqref{eq:aux2}), and the initial exponential decay is not present. This is illustrated in Fig.~\ref{figap:contour}(d), where we calculate numerically $J_{ij}^{\nu=2}$ (in solid black), and compare it with an analytical expression (in dashed red) that can be obtained in that case from the expansion of $\omega_{\nu=2}(0^+ +iy)$. This analytical expression reads:
\begin{align}
J_{ij}^{\nu=2}=-\frac{|g|^2\left[-\cos(x_{ij})\mathrm{ci}(x_{ij})+\sin(x_{ij})\left(\pi-2\mathrm{si}(x_{ij})\right)\right]}{2\pi^2 t}\,,~\label{eq:Jij2}
\end{align}
where $x_{ij}=|n_{ij}|D_l/(\pi t)$, and $\mathrm{ci}(x)$ and $\mathrm{si}(x)$ are the cosine and sine integral functions~\cite{abramowitz66a}. When $x_{ij}\ll 1$:
\begin{align}
J_{ij}^{\nu=2}\approx \frac{|g|^2\left[\gamma+\log(x_{ij})\right]}{\pi^2 t}\,,~\label{eq:Jij22}
\end{align}
with $\gamma$ being the Euler constant. This is a very interesting regime because it leads to one-dimensional photon-mediated interactions with no exponential attenuation, and with a range larger than the original photonic model.
 
$\bullet$ The case of $\nu=3$ is more difficult to treat analytically due to the logarithmic divergence that one finds in $C_3$ (see Eq.~\eqref{eq:c3}). In order to find an approximated expression, let us first note that one can approximate $G_{3}(y)$ by the lowest-order expansion of $\omega_{\nu=3}(0^+ +iy)$, that reads:
\begin{align}
G_3(y)\approx -\frac{y^2}{2\left[\left(D_l^2+y^2\left(\ln(y)-3/2\right)\right)^2+\pi^2 y^4/2\right]}\,.\label{eq:aux4}
\end{align}

By a numerical study, we showed that this approximation is already enough to reproduce well the behaviour for most distances. Then, it is convenient to divide the integrand of $J_{ij,\mathrm{BC}}^{\nu=3}$ in two functions:
\begin{align}
\label{eq:Jij3}
\frac{J_{ij,\mathrm{BC}}^{\nu=3} t}{g^2}= \int_0^\infty dy G_3(y) e^{-y|n_{ij}|}\approx \int_0^\infty dy F(y)H(y,|n_{ij}|)\,,
\end{align}
where:
\begin{align}
F(y)&= -\frac{1}{2\left[\left(D_l^2+y^2\left(\ln(y)-3/2\right)\right)^2+\pi^2 y^4/2\right]}\,, \\
H(y,d)&= y^2 e^{-y d}\,.
\end{align}

The function of $H(y,d)$ scales as
\begin{align}
H(y\ll 1,d)&\approx y^2 \\
H(y\gg 1,d)&\approx y^2 e^{-y d}\,,
\end{align}
and has a maximum $y_{h}\approx 2/d$. The function $F(y)$ scales as:
\begin{align}
F(y\ll 1)&\approx -\frac{1}{2\delta^2}\,, \\
F(y\gg 1)&\approx -\frac{1}{2 y^4 \log(2/d)^2}\,,
\end{align}
and has a maximum $y_{f}\approx \sqrt{D_l/t}/2$. Thus, it will be the ratio $y_f/y_h$ what will determine a transition between two qualitatively different regimes. For example, when $|n_{ij}|\gg 4/\sqrt{D_l/t}$, one can find the same asymptotic scaling than in the upper band-gap regime:
\begin{align}
\frac{J_{ij,\mathrm{BC}}^{\nu=3} t}{g^2}\approx -\frac{t^2}{|n_{ij}|^3 D_l^2}\,.\label{eq:jij33}
\end{align}

To find an expression for $J_{ij,\mathrm{BC}}^{\nu=3}$ in the short-distance regime,  $|n_{ij}|\ll 4/\sqrt{D_l/t}$, we found it is a good approximation to replace $\log(y)\rightarrow \log(2/d)$, which is justified due to the peaked behaviour of $H(y,d)$ around $y_h=2/d$. Like this, one can rewrite $F(y)$ as the sum of:
\begin{align}
F(y)=-\frac{|A|^2}{4 D_l^2 y^2}\left[\left(\frac{1}{y^2-A}+\mathrm{c.c.}\right)+\frac{\mathrm{Re}A}{i\mathrm{Im}A}\left(\frac{1}{y^2-A}-\mathrm{c.c.}\right)\right]\,,
\end{align}
where: 
\begin{align}
A=\frac{2 i D_l}{\pi+i(3-2\log(2/d))}\,.
\end{align}

The advantage of this rewriting is that now the integral has a closed analytical expression in terms of cosine and sine integrals:
\begin{align}
\int_0^\infty dy \frac{e^{-y d}}{y^2-A}= d q(-\sqrt{A} d)\,,
\end{align}
with $q(x)$:
\begin{align}
q(x)=\frac{2\mathrm{ci}(x)\sin(x)+\cos(x)\left(\pi-2\mathrm{si}(x)\right)}{2x}\,,
\end{align}

Thus:
\begin{align}
\frac{J_{ij,\mathrm{BC}}^{\nu=3} t}{g^2}\approx -\frac{|A|^2 d }{2 D_l^2}\left[\mathrm{Re} [q(-\sqrt{A}|n_{ij}|)]+\frac{\mathrm{Re}A}{\mathrm{Im}A}\mathrm{Im} [q(-\sqrt{A}|n_{ij}|)]\right]\,. \label{eq:jij333}
\end{align}

\begin{figure}
	\centering
	\includegraphics[width=\linewidth]{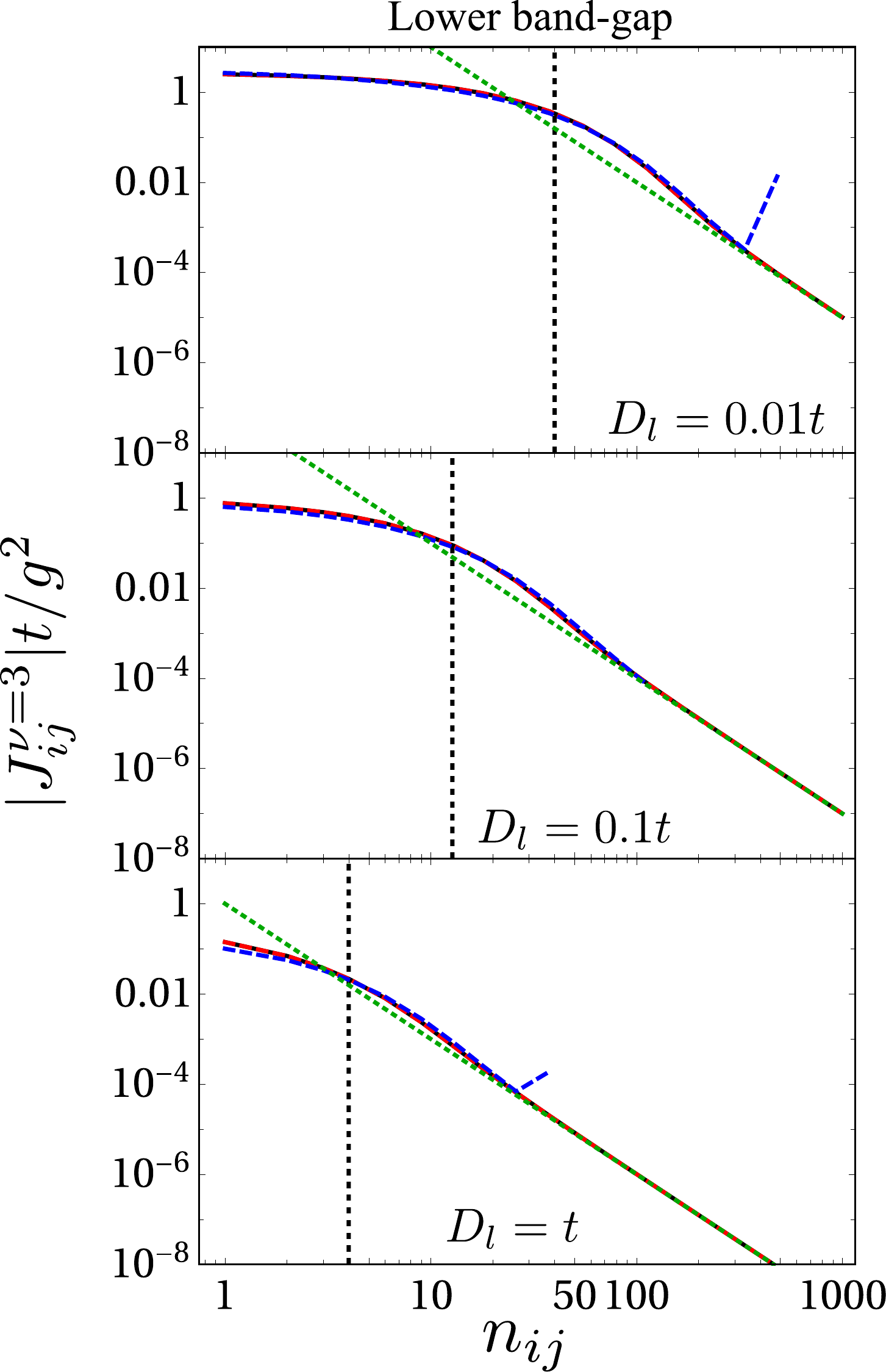}
	\caption{$|J_{ij}^{\nu=3}|$ calculated from numerically integrating Eq.~\eqref{eq:jijmu} for an emitter tuned to the lower-band gap (solid black). In dashed red, we plot the result of approximating the integrand of the branch-cut contribution as in Eq.~\eqref{eq:aux4}. In dashed green, we plot the asymptotic expression of Eq.~\eqref{eq:jij33}. In dotted blue, we plot the approximated expression we found in Eq.~\eqref{eq:jij333}. The three panels correspond to increasing values of $D_l/t$ (see legend), whereas the vertical dashed line correspond to the critical distance $4/\sqrt{D_l/t}$ where the transition between the short/long-range behaviour is expected.}
	\label{figap:contour2}
\end{figure}

In Fig.~\ref{figap:contour2} we illustrate the behaviour of all these approximations. In the different panels, we compare the result of numerically integrating Eq.~\eqref{eq:jijmu} for $\nu=3$ and several detunings $D_l/t=0.01,0.1,1$ (in solid black) with the different approximated expressions we found. For example, in dashed red we plot the result of calculating the branch-cut detour contribution approximating the integrand by the expression given by Eq.~\eqref{eq:aux4}. In dashed green, we plot the asymptotic contribution found in Eq.~\eqref{eq:jij33}, whereas in dotted blue we plot the expression we found for short distances in Eq.~\eqref{eq:jij333}. We note that at long-distances it starts to deviate significantly due to the approximation $\log(y)\rightarrow \log(2/d)$ we perform to be able to obtain an analytical expression. However, this occurs when the asymptotic expression of Eq.~\eqref{eq:jij33} already capture well the results.

\bibliographystyle{apsrev4-1}
\bibliography{Scigood,books}

\end{document}